%% file: ourwork.tex
\UseRawInputEncoding
\documentclass[usenatbib,useAMS,usedcolumn]{mnras}

\usepackage{epsfig}
\usepackage{ulem}
\usepackage{times,pdflscape}
\pdfoutput=1
\hypersetup{pdfauthor={H.L. Liu},
            bookmarksnumbered=true}

\newcommand{\cm}{\ensuremath{\mbox{~cm}}}

\newcommand{\pcmcu}{\ensuremath{\cm^{-3}}}

\newcommand{\vel}{km\,s$^{-1}$}

\newcommand{\msun}{$M_{\odot}$}
\newcommand{\lsun}{$L_{\odot}$}
\newcommand{\um}{$\mu$m}

\newcommand{\egcite}{\citep[e.g.,][]}

\newcommand{\htcop}{H$^{13}$CO$^{+}$}

\newcommand{\filname}{G34}

\graphicspath{{./0figs/}}

\begin{document}

\title[Multi-scale structures and gas kinematics in G34
] 
{ATOMS: ALMA Three-millimeter Observations of Massive Star-forming regions---IX. A pilot study towards IRDC G034.43+00.24 on multi-scale structures and gas kinematics
}


\input{authors.txt}

\date{Accepted 03 February 2022; Received 18 January; in original 03 December 2021}

\pagerange{\pageref{firstpage}--\pageref{lastpage}} \pubyear{2022}

\maketitle 

\label{firstpage}

\begin{abstract} 
 We present a comprehensive study of the gas kinematics associated with density structures at different spatial scales in the filamentary infrared dark cloud, G034.43+00.24 (G34). This study makes use of the \htcop~(1--0) molecular line data from the {\it ALMA Three-millimeter Observations of Massive Star-forming regions} (ATOMS) survey, which has spatial and velocity resolution of $\sim 0.04$\,pc and 0.2\,\vel\ , respectively. 
Several tens of {\it dendrogram} structures have been extracted in the position-position-velocity space of \htcop, which include 21 small-scale leaves and 20 larger-scale branches. Overall, their gas motions are supersonic but they exhibit the interesting behavior where leaves tend to be less dynamically supersonic than the branches. For the larger-scale, branch structures, the observed velocity--size relation (i.e., velocity variation/dispersion versus size) are seen to follow the Larson scaling exponent while the smaller-scale, leaf structures show a systematic deviation and display a steeper slope. We argue that the origin of the observed kinematics of the branch structures is likely to be a combination of turbulence and gravity-driven ordered gas flows. In comparison, gravity-driven chaotic gas motion is likely at the level of small-scale leaf structures. The results presented in our previous paper 
and this current follow-up study suggest that the main driving mechanism for mass accretion/inflow observed in G34 varies at different spatial scales. We therefore conclude that a scale-dependent combined effect of turbulence and gravity is essential to explain the star-formation processes in G34.

\end{abstract} 

\begin{keywords}
stars: formation; ISM: clouds; ISM: kinematics and dynamics; ISM: individual objects: G034.43+00.24.
\end{keywords}

\footnotetext[1]{E-mail: hongliliu2012@gmail.com, tej@iist.ac.in, and liutie@shao.ac.cn}

\section{Introduction} \label{sec:intro}
High-mass stars ($M_{\star}>8$\,\msun) play a crucial role in the evolution of galaxies. They dictate the energy budget of the host galaxy via powerful radiation, outflows, winds, and supernova events, and are responsible for the production and transfer of heavy elements to the surrounding interstellar medium (ISM).
They also influence
future star formation in their natal molecular clouds \egcite{Ken05,Urq13}. High-mass star formation has therefore long been the subject of
intense astrophysical research after the basic scenario of isolated, low-mass star formation was established in the late 1980s \egcite{Shu87}. However, understanding the processes involved is not easy owing to the several observational challenges, e.g., large distances, short lifespans, and formation in clustered environment. Several
great efforts  
from both observations and numerical simulations \egcite{Zha09,Wan11,Per13, Wan14,Beu18,Yua18,Mot18,Vaz19,Pad20,Liu21} 
focused on the physical processes of star formation on multiple scales from molecular clouds, through filaments and clumps, and finally down to cores. These studies have revealed 
that high-mass star formation is a multi-scale, hierarchical fragmentation process that occurs on almost all relevant scales of density structures. 

Density structures at different scales are intrinsically connected both spatially and kinematically.
Spatially, they are nested in the top-down manner where
molecular clouds form on scales of several tens of parsecs with low-density molecular gas
($n \la 10^{2-3}$\,\pcmcu), evolve to clumps on  scales of several parsecs with higher densities, and finally 
 are concentrated into pre and proto-stellar cores on scales of $\sim 0.01-0.1$\,pc with even higher-density gas.
Kinematically, density structures at different scales are correlated through coherent motions. A pioneering work on turbulence in the ISM by \citet{Lar81} proposed a velocity-size scaling relation, $\delta v \propto L^{\gamma}$ with $\gamma=0.38$. 
Subsequent studies have established the universality of this scaling relation in diverse environments where a value of ${\gamma}=0.5$ is commonly seen \egcite{Sol87,Hey04}.
This suggests that the gas motions of the giant molecular clouds (GMCs), in general, can remain coherent over four orders of magnitude of scales from $\sim$100 to 0.1\,pc. Recent unprecedented, multi-wavelength Herschel observations have revealed the prevalence of intermediate-scale filamentary structures within individual molecular clouds.
With numerous detailed kinematic studies from the observational perspective 
\egcite{Hac13,Hac18,Liu19, Gon19, Alv21}, these structures have been found to be velocity coherent. 

Moreover, these observable results have been successfully reproduced with dedicated simulations \egcite{Smi14,Vaz19,Pad20,Lu 21}. For instance, recent state-of-the-art numerical simulations \egcite{Pad20} modelled the cloud complexes with a web of filaments, each with a longitudinal velocity gradient related to the directional mass flow converging towards the web node, where high-mass young stellar objects (YSOs) are preferentially formed. One of the key predictions from their simulations is that the mass of the final star is determined not only by the small clump- or core-scale mass accretion  but also by the larger-scale, filamentary mass inflow/accretion. 
In other words, the mass of the newly formed stars is strongly related to the multi-scale, coherently dynamical mass inflow/accretion (a manifestation of gas motions), which agrees well with the results from recent dedicated multi-scale kinematic observations \citep{Zha11,Per13,Yua18,Liu21}. In recent years, excellent facilities like the Submillimeter Array (SMA) and the Atacama Large Millimeter/submillimeter Array (ALMA), have opened up the possibility of studying the kinematics of density structures in high-mass star formation regions across multiple scales from clouds down to seeds of star formation thereby giving tremendous impetus to the field of high-mass star formation.

\begin{figure}
\centering
\includegraphics[width=3.0 in]{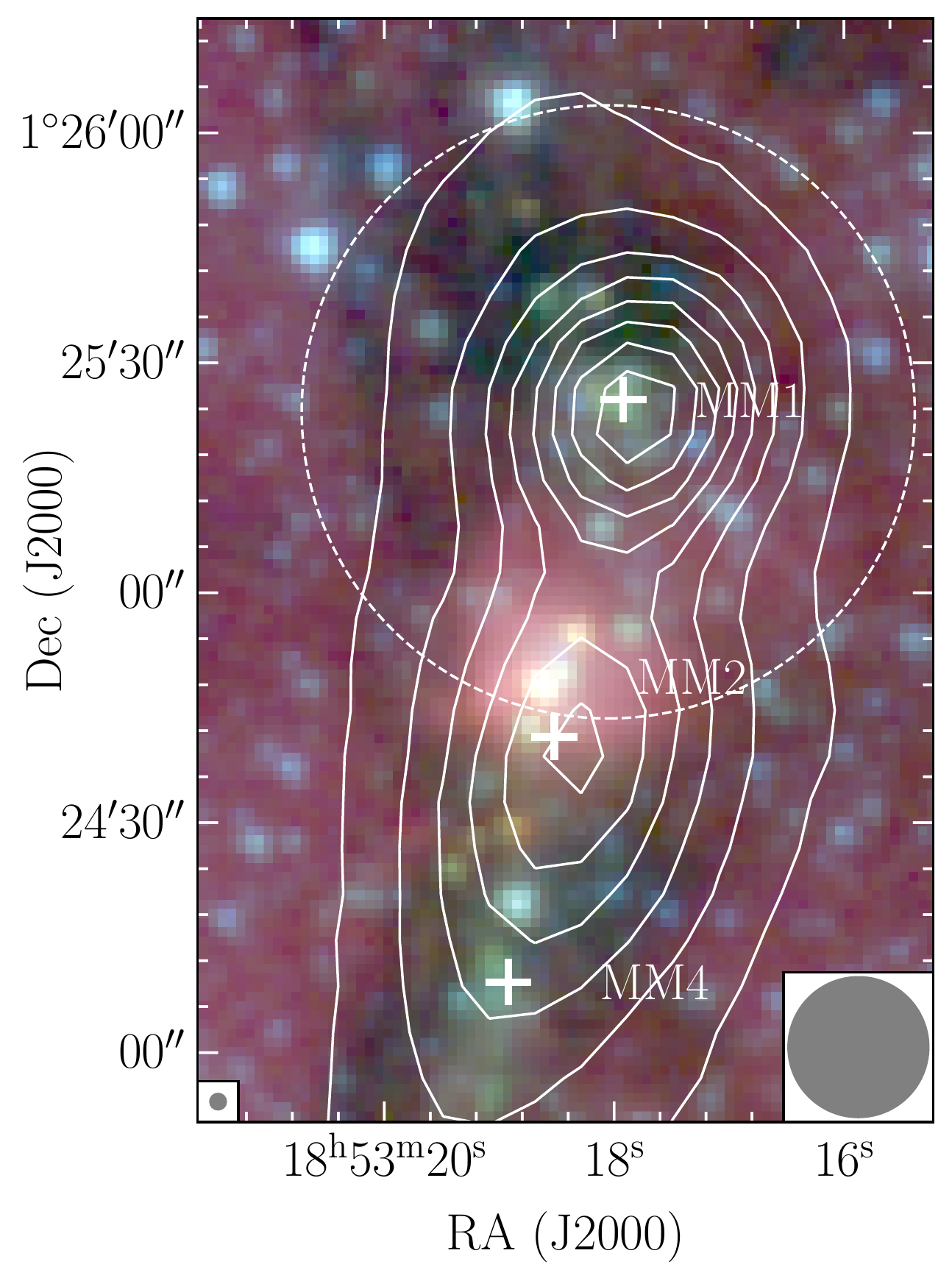}
\caption{Overview of the densest central region of G34 harbouring a massive, high density structure with ongoing star formation that is probed by ATOMS (field of view indicated by the
white dashed circle). Spitzer colour-composite image, 3.6 (blue), 4.5 (green), and 5.6\,\um\ (red) is shown superimposed with the ATLASGAL 870\,$\mu$m continuum contours. The contours start at 3\,rms (rms~$\sim0.6$~Jy beam$^{-1}$) with increasing steps defined as the power-law $D=3\times N^{p}+2$ with the dynamical range of the intensity map (i.e., the ratio between the peak and the rms noise) $D$, and the number of contours $N$ (8 in this case). The region contains three massive protostellar clumps MM1, MM2, and MM4 catalogued in \citet{Rat06}. The field of view of the ATOMS observation 
covers the clump MM1 and part of clump MM2. The beams of the Spitzer and ATLASGAL data are shown at the bottom left, and bottom right corners, respectively.}
\label{fig:overview_rgb}
\end{figure}

The target of this work is the densest central region of the filamentary cloud, G034.43+00.24 (hereafter G34, \citealt{She04,She07,Rat05,Rat06,Cor08,Liu20a, Liu21}) which is shown in Fig.\,\ref{fig:overview_rgb}. Located at $3.7\pm 0.3$\,kpc \egcite{Rat05, Tan19, Liu20a, Liu21}, \filname\ is a well-known high-mass star-forming, filamentary infrared dark cloud (IRDC) with the two  massive protostellar clumps (i.e., MM1, and MM2 in Fig.\,\ref{fig:overview_rgb}) located at the centre. These protostellar clumps have masses of the order of few hundred solar masses, sizes $\sim$0.2--0.5\,pc, and luminosities of $\sim10^4$\,\lsun\ which are typical for high-mass protostars.
Along with the IRDC nature of \filname, the massive and luminous characteristics of the two clumps reflect early stages of high-mass star formation. 

In this paper, the third in the series of studies on IRDC G34 (\citealt{Liu20a,Liu21}, hereafter Paper\,I, and Paper\,II, respectively)  and also a pilot study of multi-scale structures and kinematics using 
interferometric data from the ATOMS survey\footnote{ATOMS: ALMA Three-millimeter Observations of Massive Star-forming regions survey} (Project ID: 2019.1.00685.S,  \citealt{Liu20b,Liu20c,Liu21}, see Sect.\,\ref{sec:observations}), we 
aim to investigate the kinematics of density structures at different scales  through  the exemplary G34 cloud.
In Paper\,I, we analysed in great detail the chemical components of nine protostellar clumps (i.e., MM1--MM9, see Fig.\,1 of Paper\,I) of G34, with observations of several $\sim1$\,mm lines by the Atacama Pathfinder EXperiment telescope. The study showed that the chemistry of the clumps is closely connected to their underlying physics related to star-formation activity (e.g., luminosity and outflows). Furthermore, fragmentation and dynamical mass accretion processes during the early stages have been revealed and assessed in great detail across multiple scales in Paper\,II. The analysis carried out in Paper\,II concluded that high-mass star formation in \filname\ could be proceeding through a dynamical mass inflow/accretion process linked to the multi-scale fragments from the clouds, through clumps and cores, down to seeds of star formation.
In this work, we carry out a follow-up study 
of Paper\,II using the \htcop~(1--0) line data. The sections are organized as follows: Section\,2 briefly describes the ALMA-ATOMS data, 
Section\,3 presents the dendrogram analysis of the \htcop~(1--0) data including the identification of dendrogram structures, and their
 morphological and kinematic analysis, Section\,4 discusses the cloud dynamical state,  Section\,5 gives its implication on star formation, and Section\,6  summarizes the results.

\section{ALMA data}
\label{sec:observations}
The MM1 and MM2 clumps of G34 are covered in the observations of 
the target source I18507+0121 from the ATOMS survey. The detailed descriptions about the scientific goals of the survey, observing setups and data reduction can be found in \citet{Liu20b,Liu20c,Liu21a} and Paper\,II. In this paper, we only make use of the 12m+ACA combined data of H$^{13}$CO$^+$~(1--0). 
This line emission displays mostly single-peak profiles across the entire observed region (see Fig.\,\ref{fig:outflow:gallary}). This is expected since the H$^{13}$CO$^+$~(1--0) line is generally considered to be relatively optically thin. In Paper\,II, we estimated the optical depth of this line towards the nine detected dense cores. The values were found to lie in the range of 0.04--0.39, thus supporting the above assumption for the G34 complex under study. 
This makes H$^{13}$CO$^+$~(1--0) emission line a good probe for tracing the kinematics and density structures at different scales of G34 without
the complex effects from the optical depth and multiple velocity components.
It is worth mentioning that H$^{13}$CN~(1--0), another optically thin molecular line, was also observed as part of the ATOMS survey. This dense gas tracer has hyperfine components where the isolated one has been used in literature to probe the gas kinematics. However, this isolated component is usually found to be too weak to trace extended emission as compared to H$^{13}$CO$^+$~(1--0) and hence not considered for the analysis presented in this work.
Briefly, the combined data have a beam size of $1.9\arcsec \times 2.1\arcsec$, which corresponds to 0.04\,pc at the distance of G34, and a sensitivity of $\sim 8$\,mJy~beam$^{-1}$ at a velocity resolution of 0.2\,\vel.

\section{Dendrogram analysis of \htcop~(1-0)}
\begin{table*}
\centering
\caption{Parameters of dendrogram structures.}
\label{tab:cores}
\resizebox{18cm}{!}{
\begin{tabular}{cccccccccccccc}
\hline\hline
\input ./0table/atoms_B3_12M_h13co_dendro_cat_hd_mnras.tbl
\hline
\input ./0table/atoms_B3_12M_h13co_dendro_cat.tbl
\hline
\end{tabular}
}

\begin{flushleft}
{\bf Note:} the size, $L$, is calculated as $\sqrt{\rm maj. \times min.}/3600 \times \pi/180\times D$ given the distance $D=3.7$\,kpc of G34.
The kinetic temperature of each structure, $T_{\rm kin}$, and its uncertainty are derived from the NH$_3$ kinematic temperature map taking the mean value and standard deviation, respectively, in the structure. The mean velocity of a structure, $\langle V_{\rm lsr} \rangle$, is computed from the intensity-weighted first moment and the uncertainty of $\langle V_{\rm lsr}\rangle$ from $\delta V_{\rm lsr}$/$\sqrt{N}$ where $N$ is the number of independent beams within the structure. The velocity variation of a structure, $\delta V_{\rm lsr}$, is inferred from the standard deviation of $V_{\rm lsr}$ in the structure and the uncertainty of $\delta V_{\rm lsr}$ from $\delta V_{\rm lsr}/\sqrt{2(N-1)}$. The mean velocity dispersion of a structure, $\langle \sigma \rangle$ is given by the intensity-weighted second moment and the uncertainty of $\langle \sigma \rangle$ by $\langle \sigma \rangle$/$\sqrt{N}$ with the number of independent beams within the structure $N$. The last column, Assoc., indicates the association between the leaves and dust continuum cores identified in \citet{Liu21}.
\end{flushleft}
\end{table*}
\subsection{Dendrogram structure identification}
\label{sec:dendro_identify}

\begin{figure*}
\centering
\includegraphics[width=6.9 in]{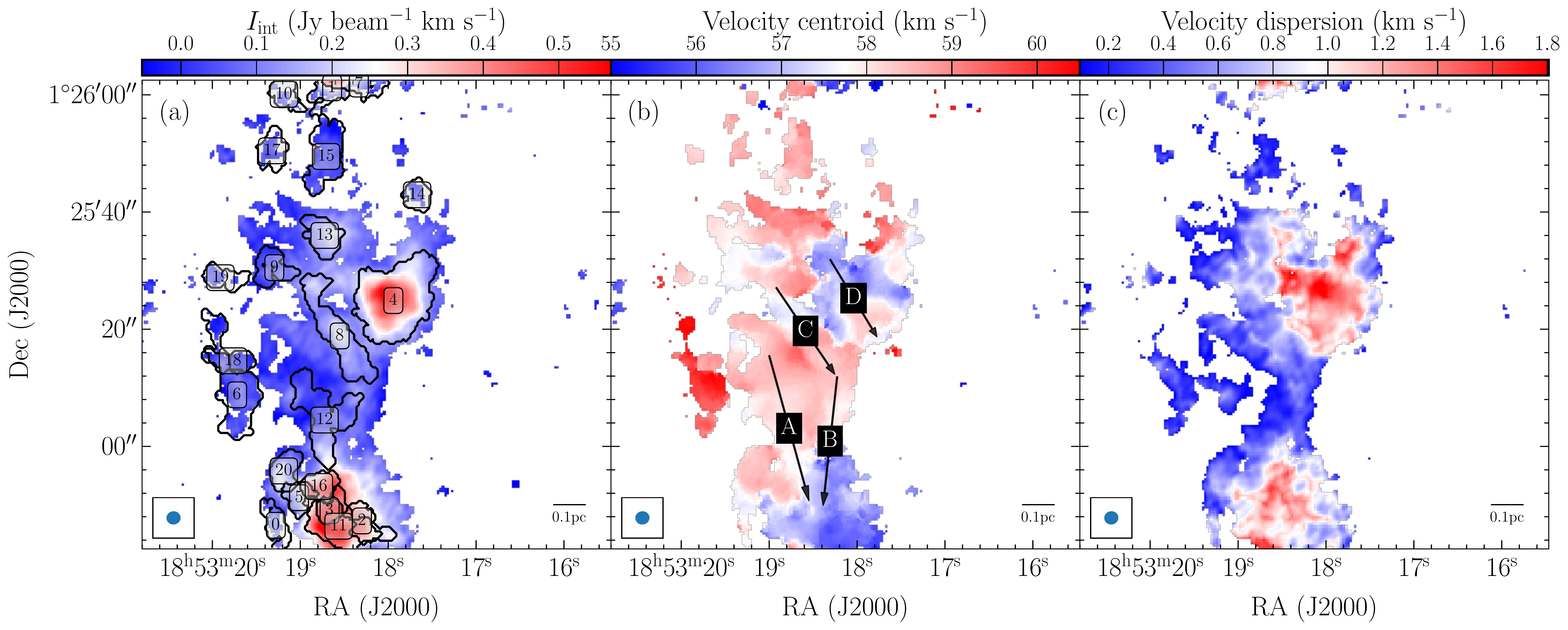}
\caption{ Moment maps of \htcop\ adapted from Paper\,II.
(a): velocity-integrated intensity map of \htcop~(1--0) for the G34 region probed with the ATOMS survey
The integrated velocity range is 12\,\vel\ centred at the systemic velocity 57.6\,\vel. The contours with the associated numbers correspond to the leaf structures identified by the {\it Dendrogram} algorithm (see text). (b): moment\,1 map of \htcop~(1--0). 
The arrows\,A--D mark the directions of the  observed velocity-coherent gradients.  (c): velocity dispersion map of \htcop~(1--0). 
In all panels, the map is displayed only at the positions where the peak intensity of the \htcop\ spectrum is $\geq5$ times 
the local noise. The beam size is shown at the bottom left corner.
}
\label{fig:kin:maps}
\end{figure*}

Dendrogram analysis was conducted to extract the density structures of different scales from the \htcop~(1-0) data in a position-position-velocity (PPV) space using the {\it Dendrogram} algorithm\footnote{\url{https://dendrograms.readthedocs.io/en/stable/index.html}}. 
Following \citet{Ros08}, the dendrogram methodology works as a tree diagram that characterises emission structures as a function of 
the level of three-dimensional intensity isocontours,
and arranges them into a tree hierarchy composed of leaves and branches.
In context of the tree hierarchy, the leaves are defined as the small-scale, bright structures at the tips of the tree that do not break up into further substructures. And branches are the larger-scale, fainter structures lower in the tree that do break up into substructures.

In order to optimise the performance of the {\it Dendrogram} algorithm,   noisy pixels, where the peak intensity along the spectral axis is less than five times the local noise level, were masked out to generate a modified data cube of \htcop. 
Here, the local noise level was considered since the actual noise distribution is not uniform in the ATOMS data due to different primary beam responses across the observed region. The local noise was estimated for each pixel of the data cube by taking the standard deviation of intensity in the line-free velocity channels spanning 20\,\vel\ in velocity. The masked area in the new cube of \htcop\ can be easily seen in the moment maps presented in Fig.\,\ref{fig:kin:maps}.

With the masked \htcop\ data cube,  we computed the dendrogram tree.
 The three key parameters that are used as an input to the algorithm are (i)  
{\it min\_value}~$=5 \sigma$ for the significance of the intensity peak of individual structures,  where
$\sigma$ (= 3.2\,mJy\,beam$^{-1}$) is the typical noise of the 
modified, masked data cube. There is an improvement of more than a factor of 2 in $\sigma$ of the masked data cube as compared to the original data cube;
(ii) {\it min\_delta}~$=2 \sigma$ for the minimum difference between two peaks for them to be considered separate
structures; and (iii) {\it min\_npix}~$=N$~pixels as a minimum number of  spatial-velocity pixels for a resolvable structure, where $N$ was set to be five times the synthesized beam area \egcite{Dua21}.
In total, the {\it Dendrogram} algorithm identifies several tens of structures. After discarding those that appear spurious structures near the edges, where the relative noise is higher,
we finally  retain 41 structures,  that includes 21 leaves and 20 branches. The {\it Dendrogram} tree of these structures is presented in Fig.\,\ref{fig:dendro:tree}.

In Fig.\,\ref{fig:kin:maps}a, the spatial distribution of the identified leaf structures is displayed on the \htcop~(1--0) velocity-integrated intensity adapted from Paper\,II.
It can be seen that these structures
correspond to bright emission of \htcop~(1--0) very well. 
Comparing with the 3\,mm dust continuum emission, where nine cores were detected (see Fig.\,1 of Paper\,II), the \htcop\ emission reveals more small-scale structures (21 leaves).
This difference can be understood since \htcop\ has much more extended emission than 3\,mm continuum (see Fig.\,2a of Paper\,II). However, the leaf structures extracted from the \htcop\ data are not resolved as well as the 3\,mm continuum cores even though both data have similar angular resolutions. For example, Leaf\,4 contains three 3\,mm continuum cores MM1-a/b/f identified in Paper\,II. All such associations between the leaves and 3\,mm continuum cores are listed in the last column of Table\,\ref{tab:cores}. Note that the poorly-resolved leaf structures do not affect the analysis carried out.

The measured parameters of all dendrogram structures are listed in Table\,\ref{tab:cores}. These include the coordinates, the major and minor axes, the position angle, the peak flux, and the velocity range in which the structures span. Note that the major and minor axes values listed in the table are corrected by a filling factor as described below. The original major and minor axes returned by the {\it Dendrogram} algorithm are the intensity-weighted second spatial moments along the two axes \citep{Ros08}. Consequently, the calculated area of the ellipse ($A_{\rm ell}$) of each structure is much smaller than the exact area ($A_{\rm ext}$) of the structure in the plane of the sky. To alleviate this, we define a filling factor, $\sqrt{A_{\rm ext}/A_{\rm ell}}$, which is multiplied to the original major and minor axes. Accordingly, the final size of the structures can be estimated as the geometric mean of the corrected axes (Col.\,6 of Table\,\ref{tab:cores}).

Next, the mean velocity $\langle V_{\rm lsr}\rangle$ of the structures was computed from the intensity-weighted first moment map (see Fig.\,\ref{fig:kin:maps}b) over 
their associated velocity span, the velocity variation $\delta V_{\rm lsr}$ is estimated from the standard deviation of $V_{\rm lsr}$ in the structures, and
the mean velocity dispersion $\langle \sigma \rangle$ is obtained from the intensity-weighted second moment (see Fig.\,\ref{fig:kin:maps}c). The kinetic temperature $T_{\rm kin}$ of each structure was estimated from NH$_{3}$ observations (at 3\,\arcsec\ resolution) toward G34 by \citet{Lu 14} and the non-thermal velocity 
dispersion was calculated as $\sqrt{\langle \sigma \rangle^2-\sigma^2_{\rm th}}$, where $\sigma_{\rm th}$ is the thermal component of the velocity dispersion (see Eq.\,2 of \citealt{Liu19}).  Note that not all of the structures are recovered in the interferometric kinetic temperature map.  For such structures, the $T_{\rm kin}$ values were  assumed to be the mean value of the entire map. In addition, NH$_{3}$ and \htcop\ may not trace the same gas even at these high spatial resolutions, and thus the NH$_{3}$-based temperature should be treated with caution.
All  the above parameters are  also listed in Table\,\ref{tab:cores}.

\subsection{Dendrogram morphological analysis}
\label{sec:mophology_analysis}
\begin{figure*}
\centering
\includegraphics[width=6.9 in]{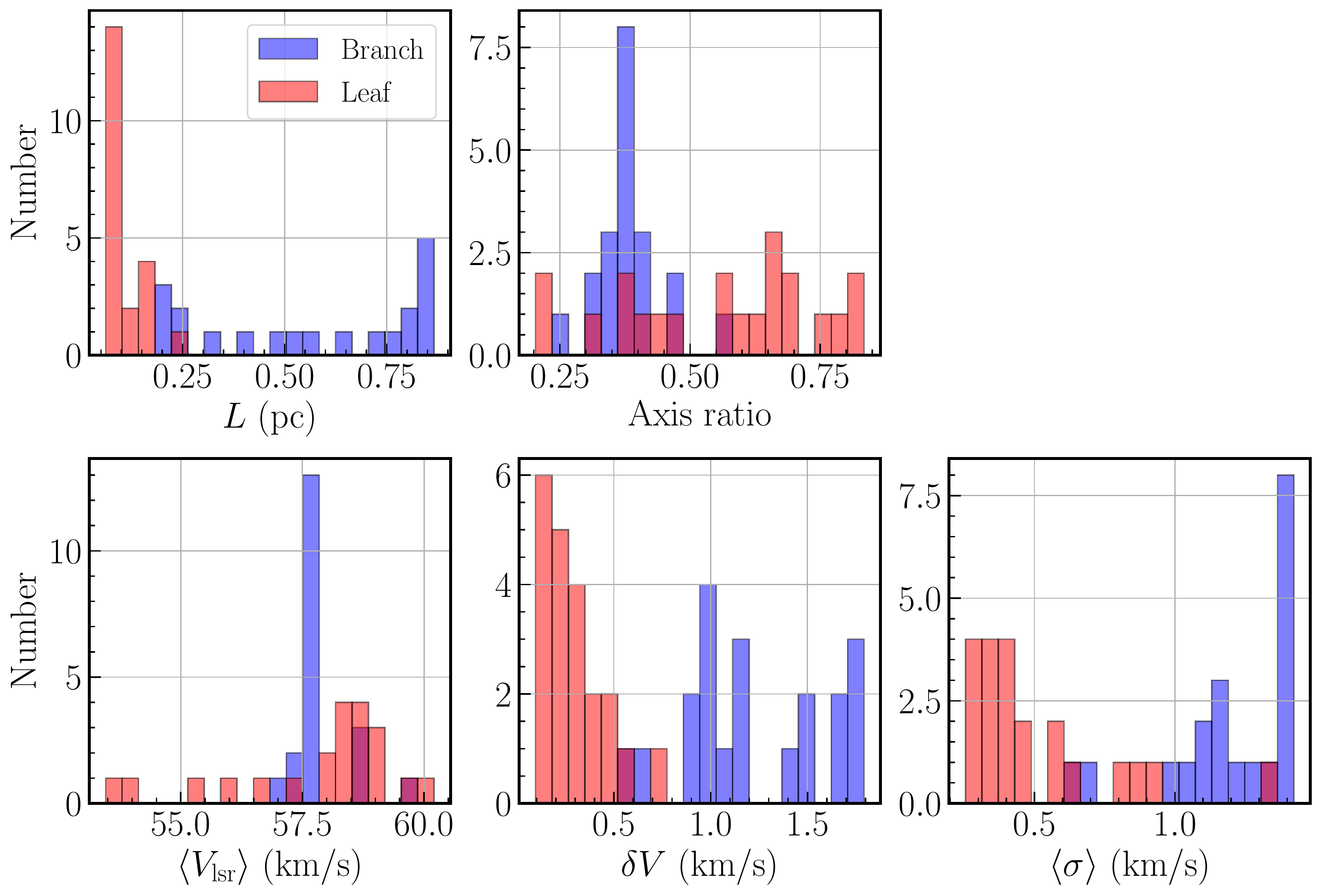}
\caption{ Histograms of leaf and branch properties: size ($L$), axis ratio, mean $V_{\rm lsr}$ ($\langle V_{\rm lsr} \rangle$), $V_{\rm lsr}$
variation ($\langle \delta V \rangle$), mean velocity dispersion ($\langle \sigma \rangle$).
}
\label{fig:dendro:hist}
\end{figure*}

Figure\,\ref{fig:dendro:hist} displays the histograms of size and axis ratio for the dendrogram structures. The size distribution
of the leaves peaks at $\sim 0.07$\,pc with a median value of 0.09\,pc within the range [0.06, 0.24]\,pc, while the branches 
 peak at $\sim 0.80$\,pc, with a median value of 0.60\,pc within [0.20, 0.87]\,pc.  There is a marginal peak seen at $\sim 0.20$ as well. It is interesting to note that size distribution of the leaves peaks at the minimum $L$ value in contrast to the branches where the peak coincides with the maximum branch size.
Overall, the typical size of the branches is greater than that of the leaves, which follows from the definition of these two different scale structures. 
Moreover, the axis ratio of the leaves show a more uniform and wider distribution with a median value of 0.59 within [0.20, 0.83], while the corresponding distribution of the branches peaks at $\sim 0.37$ with a median value of 0.37 within [0.26, 0.57]. The leaves appear, on average, more circular and are presumably more spherical compared to the branches. 
Additionally, it can be found  from the \htcop\ emission map in Fig.\,\ref{fig:kin:maps} that the peak of the axis ratio distribution of the branches nearly matches the aspect ratio ($\sim 0.35$\,pc) of the whole filamentary morphology of G34.

\subsection{Dendrogram kinematic analysis}
\label{sec:kin_analysis}
The general kinematic properties of the dendrogram structures are shown in Fig.\,\ref{fig:dendro:hist} in terms of $\langle V_{\rm lsr} \rangle$, 
$\delta V_{\rm lsr}$, and $\langle \sigma \rangle$. For the mean velocity ($\langle V_{\rm lsr} \rangle$) distribution, the branches are concentrated around the peak at the median value $\sim 57.5$\,\vel and lie within the narrow range [57.1, 59.4]\,\vel. The peak corresponds to the systemic velocity (57.6\,\vel; Paper\,II) derived from the average spectrum of \htcop\ over the entire region investigated here.
This means that majority of the branches constitute large-scale structures that contain kinematic information representative of the entire region. In contrast, the leaves  display a wide range, [53.5, 60.2]\,\vel, peaking at 58.5\,\vel\ with a median value of 58.4\,\vel. This  spread of the mean velocity for the leaves agrees with the non-uniform velocity field of G34 as shown in Fig.\,\ref{fig:kin:maps}b.

 Considering the velocity variation ($\delta V_{\rm lsr}$) distribution, we find the trend that the leaves (median 0.25\,\vel)  are on average $\sim4.5$ times lower than the branches (median 1.12\,\vel).
 Larger $\delta V_{\rm lsr}$ values for branches could be related to turbulence, gravity-driven motions, or large-scale ordered motions (see below). In particular, Leaves\,4 and 11 have $\delta V_{\rm lsr}\geq0.6$\,\vel, which fall in the overlap region with the branches. These two leaves are also among the larger ones. Here, localized outflows from star formation feedback, as revealed in Paper\,II, could be an additional source for these larger $\delta V_{\rm lsr}$ values.

For the velocity dispersion ($\langle \sigma \rangle$) distribution, the leaves peak at $\sim0.35$\,\vel\ with a median of 0.40\,\vel\ within [0.25, 1.33]\,\vel. Particularly for Leaves\,1, 4, 11, and 12, their $\langle \sigma \rangle$ values are $ \ga 0.8$\,\vel, about 2 times higher than those of the remaining leaves. 
These high values may be in part caused by feedback of ongoing star formation given the visible association of the structures and the observed star-forming activities (e.g., outflows in Leaf\,4, and  UCHII region in Leaf\,11).
In  comparison, the branches have $\langle \sigma \rangle$ values around three times on average higher than the leaves.

\section{Dynamical state of the cloud}
\label{sec:dynamic}
\begin{figure*}
\centering
\includegraphics[width=3.4 in]{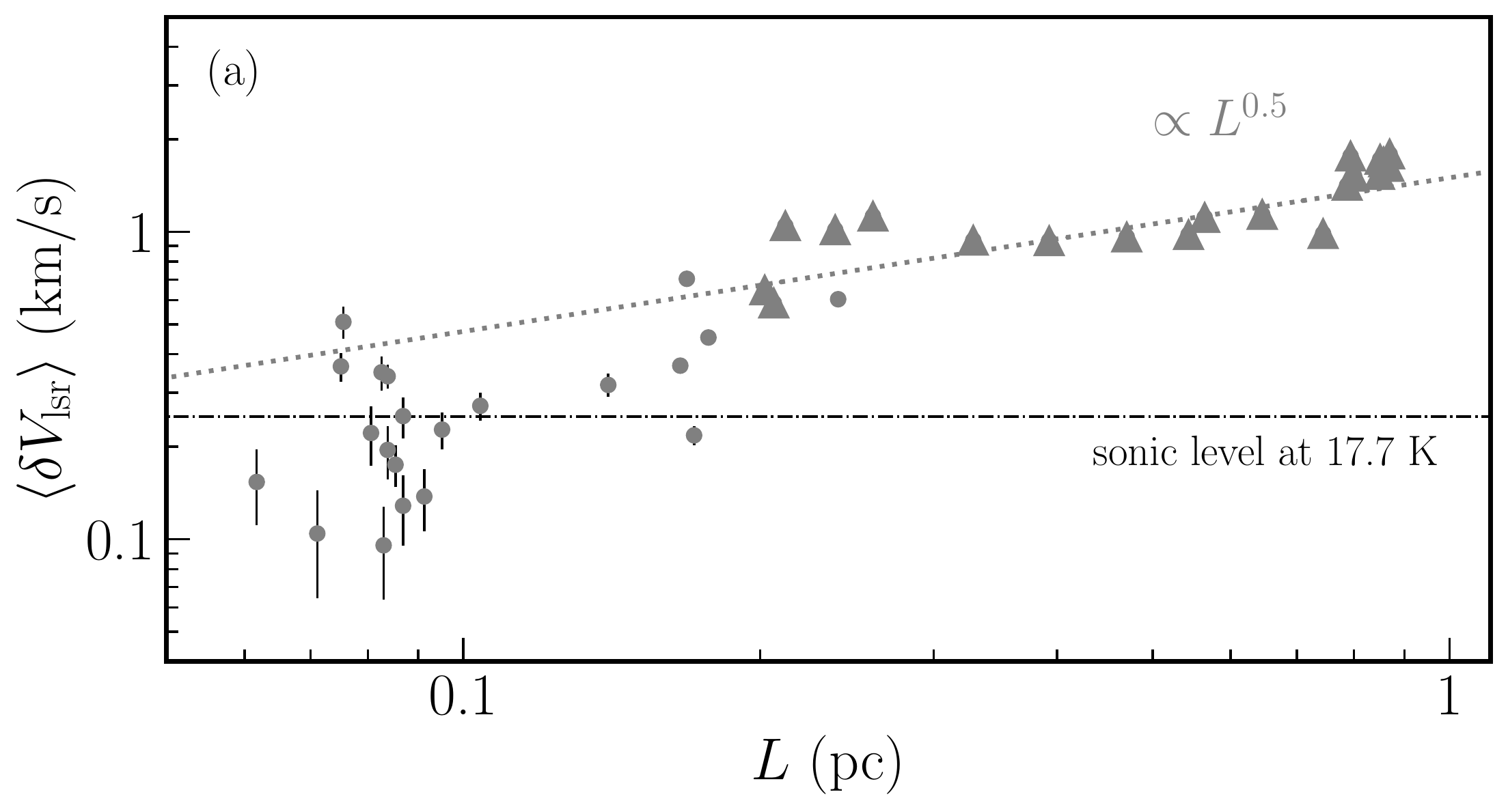}
\includegraphics[width=3.4 in]{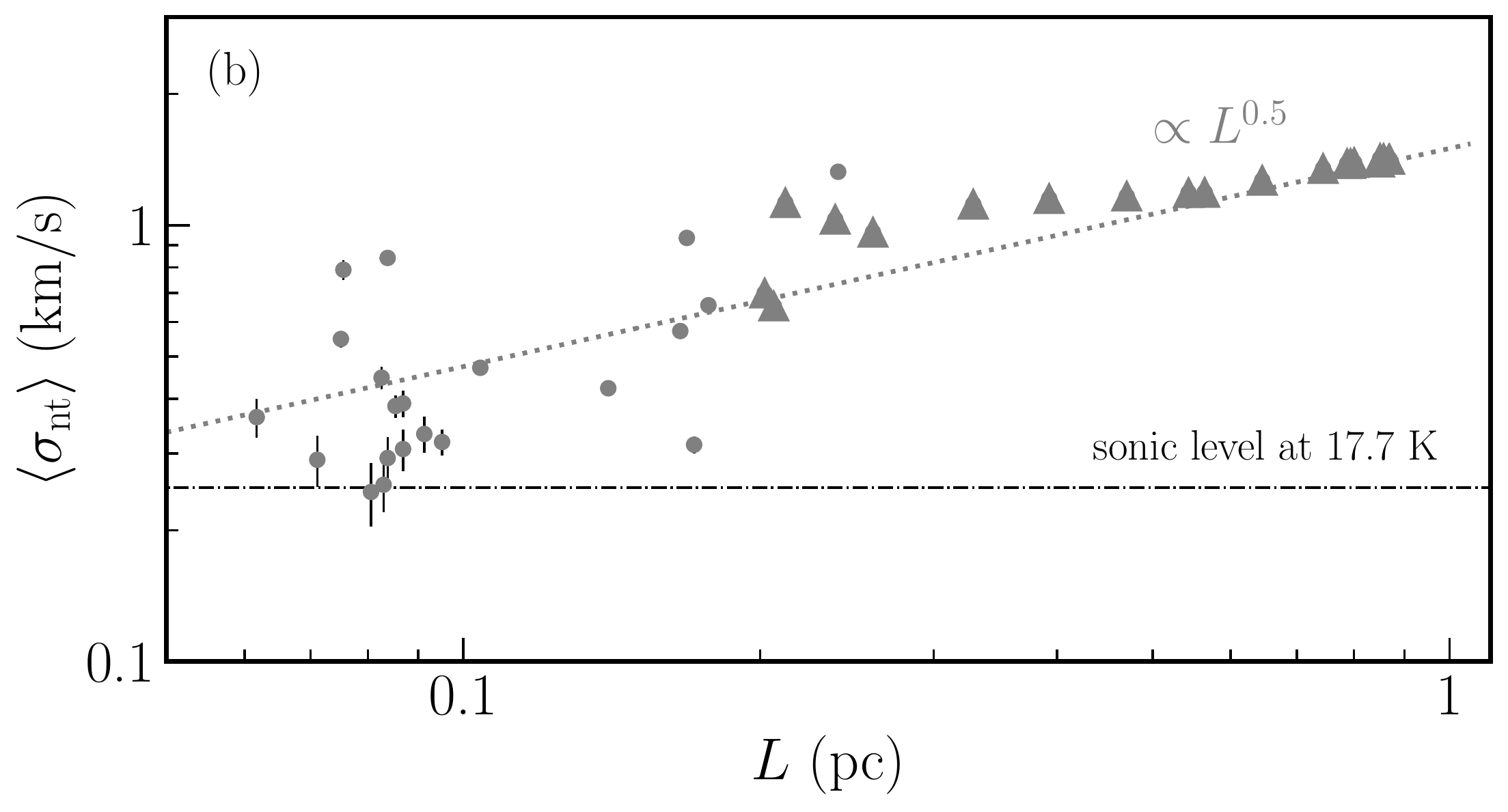}
\caption{ Velocity versus size relations: velocity variation--size relation in panel\,a and velocity dispersion--size relation in panel\,b.  Branches are shown as filled triangles and leaves as filled circles. 
The lifted Larson relation, where the coefficient is shifted from commonly quoted $\sim 1.0$ \egcite{Sol87,Hey04} to 1.5 for a clear comparison,  and the sonic level at a temperature of $\sim 17.7$\,K are indicated in dotted, and dash-dotted lines, respectively.
}
\label{fig:size-linewidth}
\end{figure*}

For an accurate
description of the cloud dynamics  one must consider the nature of the observed gas motions in the cloud. The gas motions can be  deciphered from both the velocity dispersion ($\sigma$) and the velocity variation ($\delta V_{\rm lsr}$).
 At the cloud scales, these two parameters can simply be understood to represent the mean gas motion of the entire cloud. Whereas, when we consider the substructures within individual clouds, they are distinguishable in describing different kinematics. At these smaller scales,
the parameter $\sigma$ is a measure of the mean gas motions along the line of sight of an individual structure while $\delta V$ measures the gas motions across different spatial scales on the plane of the sky inside the structure 
\egcite{Sto14,Lee14}. In  the analysis that follows, 
we investigate two scaling relations of the identified dendrogram density structures  (i.e., leaves and branches) in G34: the velocity variation versus size ($\delta V_{\rm lsr}$--$L$) and the velocity dispersion versus size ($\sigma$--$L$). 

Figure\,\ref{fig:size-linewidth} presents the  $\delta V_{\rm lsr}$--$L$ and $\sigma_{\rm nt}$--$L$ relations, where $\sigma_{\rm nt}$ is the non-thermal component of the velocity dispersion (see Sect.\,\ref{sec:dendro_identify}).
 As is evident from the plots, the dendrogram density structures have $\sigma_{\rm nt}$ above the sonic level of 0.25\,\vel, which was calculated following Eq.\,2 of \citet{Liu19} for an average kinetic temperature of 17.7\,K over the entire region investigated here. A similar trend is also seen in the values of $\delta V_{\rm lsr}$ except for the lower (smallest scale) end of the leaves. Barring these leaves, where the gas motion across these structures in the plane of the sky
 could be subsonic, the results indicate overall supersonic gas motion in the density structures of \filname.
 Several observational and simulation studies of molecular clouds suggest that the supersonic
motions can be attributed to random turbulent motions, gravity-driven chaotic motions, or even ordered motions  like localized rotation and outflows, or large-scale directional gas flows. 
Except for the  systematic motions from rotation or outflows, other types of motions can generate 
velocity-size correlations that take on power-law forms such as the  widely quoted Larson relation responsible for the turbulent motions \egcite{Lar81,Sol87,Hey04}, 
the Heyer relation $\delta V_{\rm lsr}^2/L \propto$~constant (with the mass surface density $\Sigma$) responsible for the gravity-driven chaotic motions \egcite{Hey09,Bal11,Vaz19}. 

 As shown in Fig.\,\ref{fig:size-linewidth}, the $\delta V_{\rm lsr}$--$L$ and $\sigma_{\rm nt}$--$L$ functions for the density structures identified in \filname\ can be represented by different power-laws slopes. For the larger scale structures, the branches (i.e., $>0.2$\,pc in size), the power-law dependence for both these functions can be described by scaling exponents that are in very good agreement with the predictions of \citet{Lar81,Hey04}. This implies that the universality of the ISM turbulence inferred from GMCs is valid for density structures in the range $\sim 0.2 - 1$~pc identified in \filname\ . 
 
In addition to supersonic turbulence, large-scale ordered gas motions would also produce the power-law velocity scaling. Dominance of ordered motion is seen to result in deviation from the empirical Larson's slope \citep{Hac16}. For G34, such large-scale velocity-coherent gradients are evident in Fig.\,\ref{fig:kin:maps}b along several directions, for example A and B, towards the cluster of dense, small-scale structures within MM2 (i.e., dust cores in Paper\,II).
However, for the larger-scale branch structures, the observed velocity scaling is seen to be consistent with the Larson relation. This indicates that the large-scale ordered gas motion in G34 are likely to be kinematically coupled to the turbulence cascade.
 
In contrast, the leaf structures present a different picture in the velocity scaling (see Fig.\,\ref{fig:size-linewidth}). Notwithstanding the large scatter seen, there is a clear indication of a steeper slope for the leaves in both plots. 
This systematic deviation from the empirical Larson law (i.e., $\propto L^{0.5}$) implies that the driving mechanism for gas kinematics at this level is different.

Theoretically, the gravity-driven chaotic motions,  which are the result of the hierarchical gravitational collapse (GHC) process \egcite{Vaz19}, are thought to 
explain the deviation of the velocity scaling from the empirical Larson's law \egcite{Bal11}. 
Such motions are expected to take place in local centers of collapse that appear on any scales within the cloud depending on the evolution of its density distribution.  Such collapsing centres have large thermal pressures (exceeding the mean ISM values) and develop a pseduo-virial state where rather than virial equilibrium the total energy of the system is conserved and
characterized by the power-law relation $\delta V_{\rm lsr}^{2}/L=2G\Sigma$  where $\Sigma$ is the mass surface density and G the gravitational constant.
This relation implies that massive, compact cores generally have larger velocity dispersions for larger mass surface densities (see Fig.\,2 of \citealt{Bal11}). To this end, we attempted to inspect the $\delta V_{\rm lsr}^2/L$--$\Sigma$ and $\sigma_{\rm nt}^2/L$--$\Sigma$ diagrams, where $\Sigma$ of each structure was estimated from the mean integrated intensity of the \htcop\ line emission. Here, we assume local thermodynamic equilibrium (LTE), and the abundance ratio [\htcop]/[H$_{2}$] = $2\times10^{-9}$ (Paper\,I). 
 But from these plots (not shown here) for both leaves and branches, we do not find any evidence of an increasing trend of $\delta V_{\rm lsr}^2/L$ with $\Sigma$ nor $\sigma_{\rm nt}^2/L$ with $\Sigma$, where $\Sigma$ has a dynamical range of $\sim 0.1$ to 5.0\,g~cm$^{-2}$.  For the branches, this result can be easily understood to be due to the dominance of the large-scale supersonic turbulence as mentioned earlier. However, for the small-scale leaves, the absence of the above-mentioned increasing trend could result from the insufficient dynamical range of $\Sigma$ (i.e., only about one order of magnitude in range) probed with the current data set. Hence, the dominance of gravity-driven chaotic motions at the scale of the leaves cannot be ruled out.

 Based on the different slopes
 observed and the analysis discussed above, the picture that unfolds for G34 can be understood as follows. For larger-scale, filamentary density structures (i.e., the branches), where the velocity structure is seen to be consistent with the Larson power-law, gas kinematics is driven by  randomized turbulent motions and gravity driven ordered motion. 
In comparison, for the smaller-scale, collapsing core structures (i.e., the leaves), where a deviation is seen from the Larson slope, several scenarios can be invoked. Driving sources for supersonic motions and hence the turbulent properties could differ in high-density, localized regions as presaged by \citet{Hey04}. Compared with the single-dish CO~(1--0) (the critical density  $n_{\rm crit}\sim10^{2}$\,\pcmcu\ for an excitation temperature of $T_{\rm ex}=$\,20\,K, \citealt{Eva20, Eva21}) observations for GMCs that give rise to the Larson's law, our ALMA \htcop~(1--0) ($n_{\rm crit}\sim10^{4}$\,\pcmcu\ for $T_{\rm ex}=$\,20\,K, \citealt{Shi15}) observations do trace the relatively high-density, localized substructures within the \filname\ cloud. 
A similar broken power-law velocity field was also observed in the sonic, filamentary cloud, Musca \citep{Hac16} where large-scale and ordered velocity gradients following Larson-type scaling power-law exponent are seen at scales greater then $\sim$1pc and a deviation in the form of a shallower slope is seen for smaller scales.  These authors suggest Musca to be a filament where the gas kinematics is fully decoupled from turbulence and suggest that individual regions within molecular clouds need not follow the global properties described by the Larson's relation.
A similar velocity field in G34 or the dominance of chaotic gravitational collapse cannot be ruled out at the core level though it is not decipherable with the current analysis. A rigorous study of a larger sample at high resolution is essential to confirm the steeper scaling exponent and further establish the role of various driving mechanisms of supersonic gas motion at the core scales \egcite{Yun21}. It would also help understand if internal gas kinematics of density structures within individual regions can differ from the global properties of molecular clouds.

\section{Implications for star formation}
\label{sec:imply}


In Paper\,II, we presented observational evidence of multi-scale fragmentation from clouds, through clumps and cores, down to seeds of star formation, and the cascade of scale-dependent mass inflow/accretion. From these results, which are in good agreement with the predictions of hierarchical fragmentation-based models, e.g., “global hierarchical collapse” (GHC) \citep{Vaz19} and “inertial-inflow” \citep{Pad20} models, we inferred that G34 could be undergoing a dynamical mass inflow/accretion process toward high-mass star formation. Both models propose that large-scale mass inflow rate can control the small-scale mass accretion rate onto the star(s) and predict a
decreasing trend of the mass accretion rates from large-scale clouds (e.g., $\sim 1$--10\,pc), through the filament or clump substructures (e.g., $\sim 0.1$--1\,pc), to small-scale cores ($0.01$--$0.1$\,pc) feeding protostars. This agrees well with the mass accretion/inflow rates observed in G34 (see Paper\,II). Other recent observational results \egcite{Yua18} are also in good agreement.

The detailed analysis of the gas kinematics carried out in the present follow-up study offers an additional insight into the ongoing star formation processes in G34. Both randomized turbulent motion and gravity-driven ordered gas flow are seen to explain the supersonic gas motion in the identified branches (i.e., $>0.2$\,pc in size). With the velocity-scaling following the universal Larson's relation, it is likely that turbulence dominates over the kinematically coupled large-scale, ordered motion in these larger-scale density structures. At smaller-scales (i.e., for leaves $<0.2$\,pc), where the gas motions deviate from the Larson's scaling, one could be witnessing gravity-driven chaotic motions as predicted in the GHC model. This corroborates with the fact that 
hierarchical and chaotic gravitational cascade is known to dominate at later stages in the collapsing cores \citep{Bal11} leading to chaotic density and velocity fields. This ensues because of the non-linear density fluctuations that develop due to the initial larger-scale turbulence. Such complex velocity structures, though random, have a dominant gravity-driven mode resulting in gas motions that differ from the large-scale turbulence-driven flows. The above analysis indicates that scale-dependent driving mechanisms need to be invoked to explain the observed supersonic gas motion in G34 star-forming complex.

\section{Summary and conclusions} \label{sec:summary}
We have carried out  a pilot study of multi-scale structures and kinematics in star-forming complexes. In this study, detailed analysis has been implemented on the link 
between kinematics of density structures at different scales towards
the central dense region of G34 that harbours two massive protostellar clumps. Using the interferometric 
\htcop~(1--0) data from the ATOMS survey, several tens of dendrogram structures are identified in the position-position-velocity space, including 21 leaves and 20 branches.
They have different morphological properties with the former having smaller geometrical size and greater axis ratio than the latter.  The typical sizes of the leaves and branches are $\sim 0.09$ and $\sim 0.6$\,pc, respectively.
The gas motions in these two types of structures are overall supersonic though the leaves tend to be less supersonic
than the branches. Detailed analysis of the velocity--size relation (i.e., velocity variation versus size, and velocity dispersion versus size) and comparison with the standard Larson-type velocity-scaling show that the dominant driving mechanism for branches and leaves are different. The gas kinematics in the large-scale branch structures is consistent with large-scale, ordered gas flows that is coupled with the dominant turbulent velocity structure. Whereas, a clear systematic deviation is seen at the leaf scale, which could be attributed to gravity-driven chaotic collapse. If we consider the two state-of-the-art multi-scale fragmentation-based models, ``inertial-inflow" and GHC, at larger-scales (i.e., clouds and filaments), the former strongly favours turbulence-driven mass inflow while the latter advocates for a gravity-driven hierarchical collapse process. However, both models agree on gravity-driven mass-accretion on small scales (e.g., cores and proto-stars). Thus, based on the observed gas motions in the present study and the scale-dependent dynamical mass inflow/accretion scenario leading to high-mass star formation presented in Paper\,II, we suggest that a scale-dependent combined effect of turbulence and gravity is essential to explain the star-formation processes in G34.

\medskip
\noindent{\textbf{Acknowledgements}}\\
We thank the anonymous referee for comments and suggestions that helped improve the paper.
H.-L. Liu is supported by National Natural Science Foundation of China (NSFC) through the grant No.12103045.
T. Liu acknowledges the supports by NSFC through grants No.12073061 and No.12122307, the international partnership program of Chinese Academy of Sciences through grant No.114231KYSB20200009, and Shanghai Pujiang Program 20PJ1415500.
This research was carried out in part at the Jet Propulsion Laboratory, which is operated by the California Institute of Technology under a contract with the National Aeronautics and Space Administration
(80NM0018D0004).
S.-L. Qin is supported by NSFC under No.12033005.
L. Bronfman and A. Stutz gratefully acknowledges support from ANID BASAL project FB210003.
LB gratefully acknowledges support by the ANID BASAL project ACE210002. A. Stutz gratefully acknowledges funding support through Fondecyt Regular (project code 1180350). 
C.W.L. is supported by Basic Science Research Program through the National Research Foundation of Korea (NRF)
funded by the Ministry of Education, Science and Technology
(NRF-2019R1A2C1010851).
This paper makes use of the following ALMA data: ADS/JAO.ALMA\#2019.1.00685.S. ALMA is a partnership of ESO (representing its member states), NSF (USA), 
and NINS (Japan), together with NRC (Canada), MOST and ASIAA (Taiwan), and KASI (Republic of Korea), in cooperation with the Republic of Chile. The Joint 
ALMA Observatory is operated by ESO, AUI/NRAO, and NAOJ.
This research made use of astrodendro, a Python package to compute dendrograms of Astronomical data ({\url{http://www.dendrograms.org/}}).
This research made use of Astropy,
a community-developed core Python package for Astronomy (Astropy
Collaboration, 2018).

\noindent{\textbf{Data availability}}\\
The data underlying this article will be shared on reasonable request
to the corresponding author.

\vspace{-5mm}


\input ourwork.bbl
%
\appendix
\section{Complementary figures}

\begin{figure*}
\centering
\includegraphics[width=6.8 in]{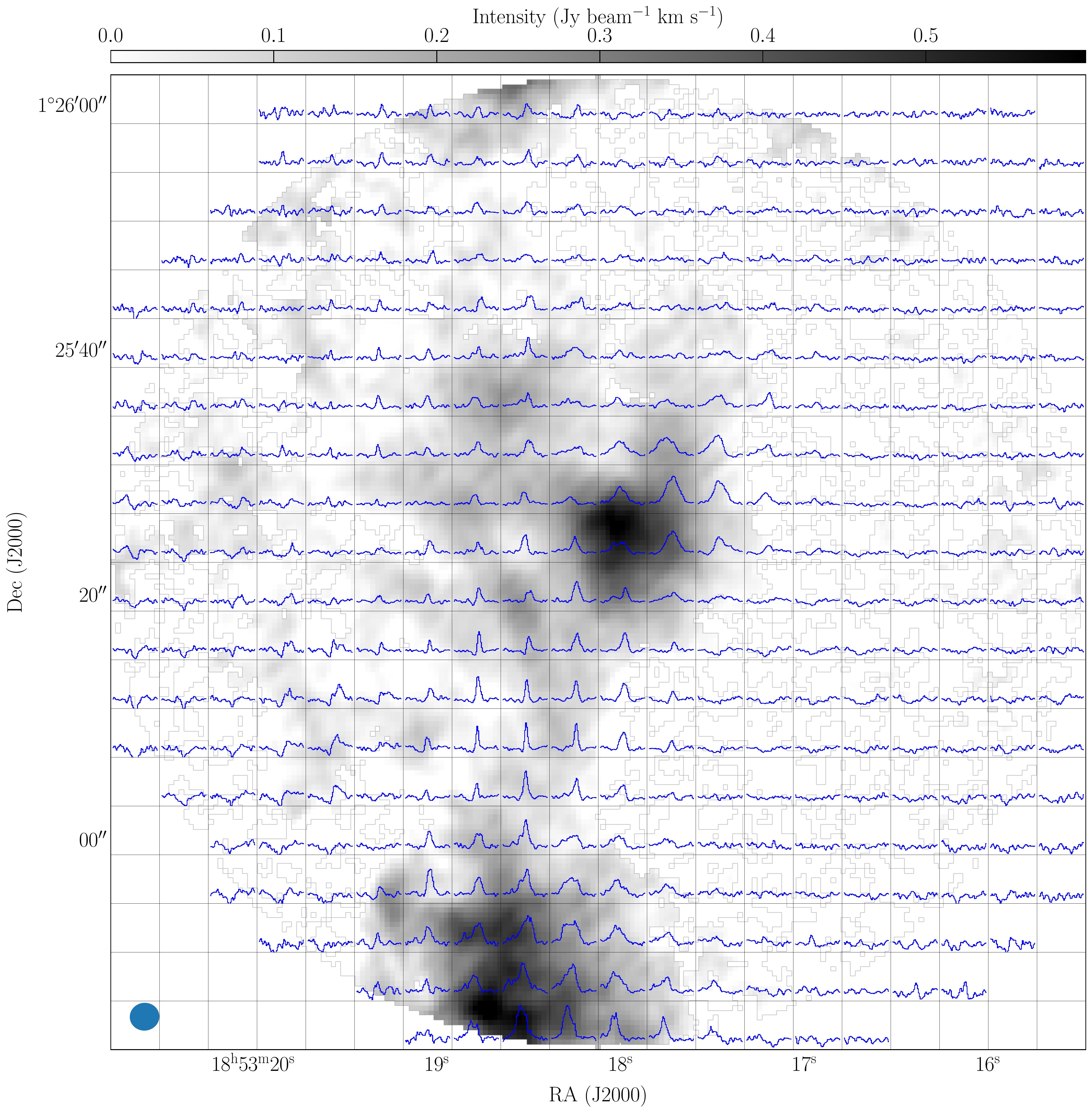}
\caption{Grid of spectra of \htcop~(1--0) overlaid on its velocity integrated intensity map for G34. The beam is displayed at the bottom left corner.
Across the entire region, except for a small fraction of areas in the MM2 clump, shows a single-peak profile of \htcop~(1--0) emission.
}
\label{fig:outflow:gallary}
\end{figure*}

\begin{figure*}
\centering
\includegraphics[width=6.9 in]{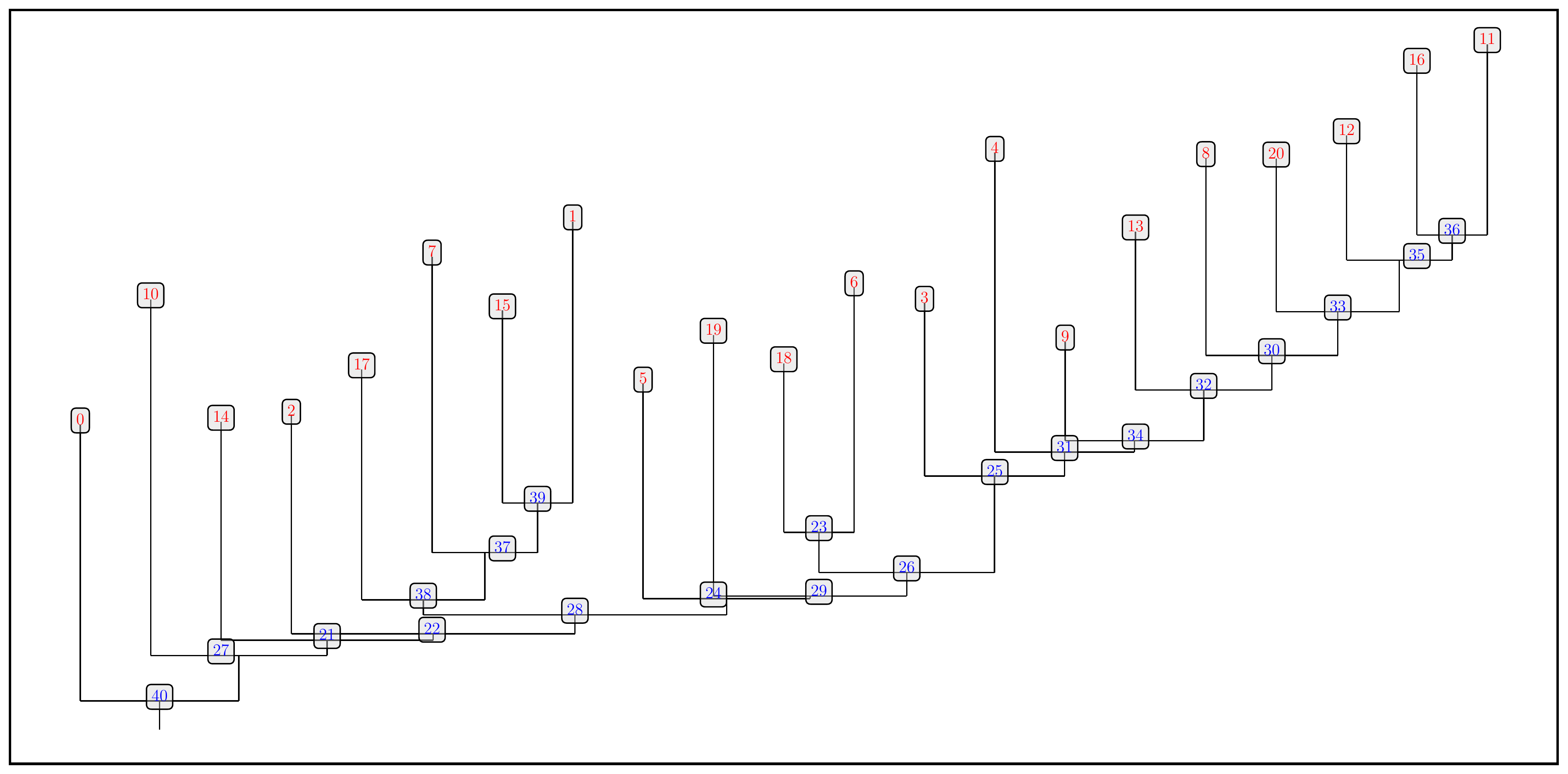}
\caption{ Dendrogram of the hierarchical structures extracted from \htcop~(1--0) emission in G34. There are 41 structures including 21 leaves (labelled in red) and 20 branches (labelled in blue). 
}
\label{fig:dendro:tree}
\end{figure*}

\input{affiliations.txt}
\label{lastpage}
\end{document}

%% file: authors.txt
\author[H.-L. Liu et al.]{
Hong-Li Liu,$^{\star 1}$
Anandmayee Tej,$^{2}$
Tie Liu,$^{3,4}$
Paul F. Goldsmith,$^{5}$
Amelia Stutz,$^{6,7}$
Mika Juvela,$^{8}$
\newauthor
Sheng-Li Qin,$^{1}$
Feng-Wei Xu,$^{9,10}$
Leonardo Bronfman,$^{11}$
Neal J. Evans,$^{12,13}$
Anindya Saha,$^{2}$
\newauthor
Namitha Issac,$^{14}$
Ken'ichi Tatematsu,$^{15}$
Ke Wang,$^{9,10}$
Shanghuo Li,$^{16}$
Siju Zhang,$^{9}$
\newauthor
Tapas Baug,$^{17}$
Lokesh Dewangan,$^{18}$
Yue-Fang Wu,$^{10,9}$
Yong Zhang,$^{19}$
Chang Won Lee,$^{16,20}$
\newauthor
Xun-Chuan Liu,$^{3,4}$
Jianwen Zhou,$^{21}$
Archana Soam,$^{22}$
\\
Affiliations are listed at the end of the paper}

%% file: 0table/atoms_B3_12M_h13co_dendro_cat_hd_mnras.tbl
    ID &            $RA$  &          $Dec.$ & ${\rm maj \times min}$ axis  &              PA &             $L$ &         $F^{p}$ &   $T_{\rm kin}$ &      Vel. range  & $\langle V_{\rm lsr}\rangle$  &      $\delta V$ & $\langle\sigma\rangle$ & $\langle\sigma_{\rm nt}\rangle$ & Assoc. \\
       &         (J2000)  &         (J2000) & ${\rm \arcsec  \times \arcsec}$  &             deg &              pc & mJy beam$^{-1}$ &               K &     km s$^{-1}$  &     km s$^{-1}$  &     km s$^{-1}$ &     km s$^{-1}$ &     km s$^{-1}$   &  \\
\hline 

\multicolumn{14}{c}{\bf Leaves} \\

%% file: 0table/atoms_B3_12M_h13co_dendro_cat.tbl
      0  &      18:53:19.28  &      +1:24:46.53  &     8.4  $\times$     3.2  &   111.1  &    0.08  &    63.8  &     9.4  $\pm$     7.5  & [   59.57  ,   61.26  ] &   60.21  $\pm$    0.06  &    0.35  $\pm$    0.04  &    0.45  $\pm$    0.02  &    0.45  $\pm$    0.03  &   \\ 
      1  &      18:53:18.64  &      +1:26:01.19  &     6.3  $\times$     4.3  &   201.4  &    0.08  &   130.5  &    17.7  $\pm$    10.8  & [   57.46  ,   60.63  ] &   58.98  $\pm$    0.04  &    0.34  $\pm$    0.03  &    0.84  $\pm$    0.02  &    0.84  $\pm$    0.02  & MM1-g  \\ 
      2  &      18:53:18.30  &      +1:24:47.29  &     7.6  $\times$     3.7  &   113.4  &    0.08  &    65.8  &    22.3  $\pm$     4.0  & [   52.82  ,   54.09  ] &   53.46  $\pm$    0.04  &    0.17  $\pm$    0.03  &    0.39  $\pm$    0.02  &    0.39  $\pm$    0.02  &   \\ 
      3  &      18:53:18.67  &      +1:24:49.05  &     4.6  $\times$     3.7  &    50.3  &    0.06  &    98.0  &    28.6  $\pm$     3.0  & [   53.46  ,   54.72  ] &   54.12  $\pm$    0.06  &    0.15  $\pm$    0.04  &    0.37  $\pm$    0.03  &    0.36  $\pm$    0.04  & MM2-a  \\ 
      4  &      18:53:17.94  &      +1:25:24.89  &    15.3  $\times$    12.1  &    49.8  &    0.24  &   166.1  &    24.0  $\pm$     3.0  & [   54.30  ,   60.42  ] &   57.46  $\pm$    0.02  &    0.60  $\pm$    0.01  &    1.33  $\pm$    0.01  &    1.33  $\pm$    0.01  & MM1-a/b/f  \\ 
      5  &      18:53:19.01  &      +1:24:51.25  &     5.7  $\times$     4.7  &   115.6  &    0.08  &    73.7  &    11.2  $\pm$     8.2  & [   54.72  ,   55.57  ] &   55.19  $\pm$    0.04  &    0.10  $\pm$    0.03  &    0.26  $\pm$    0.03  &    0.25  $\pm$    0.03  &   \\ 
      6  &      18:53:19.72  &      +1:25:08.80  &    16.8  $\times$     6.1  &   101.5  &    0.18  &   103.5  &    17.7  $\pm$    10.8  & [   58.73  ,   61.05  ] &   59.82  $\pm$    0.03  &    0.45  $\pm$    0.02  &    0.66  $\pm$    0.01  &    0.66  $\pm$    0.01  &   \\ 
      7  &      18:53:18.33  &      +1:26:01.83  &    10.3  $\times$     2.2  &   177.2  &    0.07  &   115.3  &    17.7  $\pm$    10.8  & [   55.14  ,   57.46  ] &   56.03  $\pm$    0.05  &    0.37  $\pm$    0.04  &    0.55  $\pm$    0.02  &    0.55  $\pm$    0.02  &   \\ 
      8  &      18:53:18.55  &      +1:25:18.86  &    21.2  $\times$     4.3  &   120.5  &    0.17  &   163.0  &    12.9  $\pm$     2.3  & [   57.25  ,   59.36  ] &   58.32  $\pm$    0.02  &    0.37  $\pm$    0.01  &    0.58  $\pm$    0.01  &    0.57  $\pm$    0.01  & MM1-c/d/e  \\ 
      9  &      18:53:19.30  &      +1:25:30.52  &     6.7  $\times$     5.0  &   148.8  &    0.09  &    85.5  &     3.6  $\pm$     0.1  & [   57.46  ,   58.52  ] &   57.90  $\pm$    0.04  &    0.23  $\pm$    0.03  &    0.32  $\pm$    0.02  &    0.32  $\pm$    0.02  &   \\ 
     10  &      18:53:19.19  &      +1:26:00.08  &     6.0  $\times$     3.9  &   130.6  &    0.08  &    99.2  &    17.7  $\pm$    10.8  & [   57.46  ,   60.21  ] &   58.79  $\pm$    0.09  &    0.51  $\pm$    0.06  &    0.79  $\pm$    0.04  &    0.79  $\pm$    0.04  &   \\ 
     11  &      18:53:18.56  &      +1:24:46.39  &    12.9  $\times$     7.2  &   163.3  &    0.17  &   243.6  &    22.1  $\pm$     8.2  & [   55.14  ,   58.52  ] &   56.82  $\pm$    0.03  &    0.70  $\pm$    0.02  &    0.94  $\pm$    0.01  &    0.94  $\pm$    0.01  & MM2-b  \\ 
     12  &      18:53:18.72  &      +1:25:04.50  &    12.8  $\times$     7.6  &   104.8  &    0.17  &   176.7  &    15.2  $\pm$     8.2  & [   57.68  ,   58.73  ] &   58.23  $\pm$    0.02  &    0.22  $\pm$    0.02  &    0.32  $\pm$    0.01  &    0.31  $\pm$    0.01  &   \\ 
     13  &      18:53:18.72  &      +1:25:36.01  &     7.7  $\times$     5.1  &   123.5  &    0.10  &   126.0  &    19.5  $\pm$     3.6  & [   57.89  ,   59.57  ] &   58.71  $\pm$    0.04  &    0.27  $\pm$    0.03  &    0.48  $\pm$    0.02  &    0.47  $\pm$    0.02  &   \\ 
     14  &      18:53:17.67  &      +1:25:42.93  &     5.6  $\times$     3.8  &   113.5  &    0.07  &    64.5  &    17.7  $\pm$     4.3  & [   58.31  ,   59.36  ] &   58.87  $\pm$    0.06  &    0.10  $\pm$    0.04  &    0.30  $\pm$    0.04  &    0.29  $\pm$    0.04  &   \\ 
     15  &      18:53:18.71  &      +1:25:49.47  &    12.8  $\times$     5.2  &    78.0  &    0.14  &    95.4  &     8.9  $\pm$     5.4  & [   57.68  ,   59.15  ] &   58.48  $\pm$    0.04  &    0.32  $\pm$    0.03  &    0.43  $\pm$    0.02  &    0.42  $\pm$    0.02  &   \\ 
     16  &      18:53:18.79  &      +1:24:53.16  &     6.4  $\times$     4.5  &   200.2  &    0.09  &   226.4  &    18.8  $\pm$     6.4  & [   58.10  ,   59.36  ] &   58.67  $\pm$    0.05  &    0.25  $\pm$    0.04  &    0.40  $\pm$    0.03  &    0.39  $\pm$    0.03  &   \\ 
     17  &      18:53:19.32  &      +1:25:50.45  &     6.8  $\times$     4.2  &    75.1  &    0.09  &    77.5  &    17.7  $\pm$    10.8  & [   57.89  ,   59.15  ] &   58.56  $\pm$    0.05  &    0.13  $\pm$    0.03  &    0.31  $\pm$    0.03  &    0.31  $\pm$    0.03  &   \\ 
     18  &      18:53:19.76  &      +1:25:14.62  &     9.2  $\times$     3.0  &   174.7  &    0.08  &    79.2  &    17.7  $\pm$    10.8  & [   58.52  ,   59.57  ] &   58.95  $\pm$    0.05  &    0.20  $\pm$    0.04  &    0.30  $\pm$    0.03  &    0.29  $\pm$    0.03  &   \\ 
     19  &      18:53:19.91  &      +1:25:28.79  &     7.5  $\times$     3.4  &   167.8  &    0.08  &    87.6  &    17.7  $\pm$    10.8  & [   57.89  ,   58.73  ] &   58.17  $\pm$    0.07  &    0.22  $\pm$    0.05  &    0.25  $\pm$    0.03  &    0.24  $\pm$    0.04  &   \\ 
     20  &      18:53:19.19  &      +1:24:55.81  &     7.5  $\times$     4.2  &    67.9  &    0.09  &   162.8  &    17.7  $\pm$    10.8  & [   57.89  ,   58.94  ] &   58.41  $\pm$    0.04  &    0.14  $\pm$    0.03  &    0.34  $\pm$    0.03  &    0.33  $\pm$    0.03  &   \\ 
\hline 
\multicolumn{14}{c}{\bf Branches} \\ 
\hline 
     21  &      18:53:18.51  &      +1:25:10.72  &    78.5  $\times$    29.2  &    86.9  &    0.85  &   243.6  &    18.2  $\pm$     9.6  & [   52.82  ,   61.89  ] &   57.65  $\pm$    0.01  &    1.69  $\pm$    0.01  &    1.42  $\pm$    0.01  &    1.42  $\pm$    0.01  &   \\ 
     22  &      18:53:18.51  &      +1:25:10.51  &    78.0  $\times$    28.9  &    86.9  &    0.85  &   243.6  &    18.2  $\pm$     9.6  & [   52.82  ,   61.89  ] &   57.65  $\pm$    0.01  &    1.72  $\pm$    0.01  &    1.42  $\pm$    0.01  &    1.42  $\pm$    0.01  &   \\ 
     23  &      18:53:19.73  &      +1:25:09.85  &    19.3  $\times$     6.9  &   102.6  &    0.20  &   103.5  &    17.7  $\pm$    10.8  & [   58.52  ,   61.05  ] &   59.75  $\pm$    0.04  &    0.65  $\pm$    0.03  &    0.70  $\pm$    0.02  &    0.70  $\pm$    0.02  &   \\ 
     24  &      18:53:18.51  &      +1:25:07.97  &    69.1  $\times$    28.9  &    84.0  &    0.80  &   243.6  &    19.3  $\pm$     8.3  & [   53.25  ,   61.26  ] &   57.60  $\pm$    0.01  &    1.52  $\pm$    0.01  &    1.39  $\pm$    0.01  &    1.39  $\pm$    0.01  &   \\ 
     25  &      18:53:18.46  &      +1:25:07.38  &    69.5  $\times$    24.8  &    83.7  &    0.74  &   243.6  &    19.2  $\pm$     8.1  & [   53.25  ,   61.05  ] &   57.55  $\pm$    0.01  &    0.99  $\pm$    0.01  &    1.35  $\pm$    0.01  &    1.35  $\pm$    0.01  &   \\ 
     26  &      18:53:18.50  &      +1:25:07.76  &    68.7  $\times$    28.1  &    83.8  &    0.78  &   243.6  &    19.4  $\pm$     8.3  & [   53.25  ,   61.05  ] &   57.60  $\pm$    0.01  &    1.42  $\pm$    0.01  &    1.39  $\pm$    0.01  &    1.38  $\pm$    0.01  &   \\ 
     27  &      18:53:18.51  &      +1:25:11.41  &    79.7  $\times$    29.5  &    87.5  &    0.86  &   243.6  &    18.2  $\pm$     9.8  & [   52.82  ,   62.53  ] &   57.65  $\pm$    0.01  &    1.63  $\pm$    0.01  &    1.42  $\pm$    0.01  &    1.42  $\pm$    0.01  &   \\ 
     28  &      18:53:18.52  &      +1:25:10.49  &    77.8  $\times$    28.9  &    86.9  &    0.85  &   243.6  &    18.2  $\pm$     9.6  & [   53.25  ,   61.68  ] &   57.67  $\pm$    0.01  &    1.54  $\pm$    0.01  &    1.40  $\pm$    0.01  &    1.40  $\pm$    0.01  &   \\ 
     29  &      18:53:18.51  &      +1:25:07.83  &    68.8  $\times$    28.6  &    84.1  &    0.79  &   243.6  &    19.3  $\pm$     8.4  & [   53.25  ,   61.05  ] &   57.60  $\pm$    0.02  &    1.77  $\pm$    0.01  &    1.39  $\pm$    0.01  &    1.38  $\pm$    0.01  &   \\ 
     30  &      18:53:18.63  &      +1:24:56.04  &    43.8  $\times$    15.8  &    94.6  &    0.47  &   243.6  &    20.2  $\pm$     9.9  & [   54.72  ,   59.78  ] &   57.52  $\pm$    0.01  &    0.96  $\pm$    0.01  &    1.17  $\pm$    0.01  &    1.17  $\pm$    0.01  &   \\ 
     31  &      18:53:18.46  &      +1:25:05.87  &    62.4  $\times$    20.9  &    81.9  &    0.64  &   243.6  &    20.8  $\pm$     7.8  & [   54.09  ,   60.42  ] &   57.54  $\pm$    0.01  &    1.14  $\pm$    0.01  &    1.27  $\pm$    0.01  &    1.27  $\pm$    0.01  &   \\ 
     32  &      18:53:18.63  &      +1:24:58.42  &    55.0  $\times$    16.8  &    93.8  &    0.54  &   243.6  &    19.3  $\pm$     9.6  & [   54.72  ,   60.00  ] &   57.59  $\pm$    0.01  &    0.98  $\pm$    0.01  &    1.19  $\pm$    0.01  &    1.19  $\pm$    0.01  &   \\ 
     33  &      18:53:18.65  &      +1:24:52.77  &    31.8  $\times$    15.3  &   102.8  &    0.39  &   243.6  &    24.2  $\pm$     7.3  & [   54.72  ,   59.78  ] &   57.49  $\pm$    0.02  &    0.94  $\pm$    0.01  &    1.15  $\pm$    0.01  &    1.15  $\pm$    0.01  &   \\ 
     34  &      18:53:18.63  &      +1:24:58.93  &    57.4  $\times$    17.4  &    94.5  &    0.56  &   243.6  &    19.1  $\pm$    10.0  & [   54.72  ,   60.21  ] &   57.60  $\pm$    0.01  &    1.11  $\pm$    0.01  &    1.19  $\pm$    0.01  &    1.19  $\pm$    0.01  &   \\ 
     35  &      18:53:18.62  &      +1:24:52.00  &    30.2  $\times$    11.3  &   100.2  &    0.33  &   243.6  &    24.0  $\pm$     7.4  & [   54.93  ,   59.57  ] &   57.42  $\pm$    0.02  &    0.94  $\pm$    0.02  &    1.12  $\pm$    0.01  &    1.12  $\pm$    0.01  &   \\ 
     36  &      18:53:18.60  &      +1:24:47.80  &    15.9  $\times$     9.1  &   132.6  &    0.21  &   243.6  &    21.5  $\pm$     8.6  & [   55.14  ,   59.57  ] &   57.12  $\pm$    0.04  &    1.05  $\pm$    0.03  &    1.13  $\pm$    0.01  &    1.13  $\pm$    0.01  &   \\ 
     37  &      18:53:18.66  &      +1:25:55.49  &    22.3  $\times$     8.2  &    74.7  &    0.24  &   130.5  &     5.4  $\pm$     1.9  & [   55.14  ,   61.68  ] &   58.70  $\pm$    0.03  &    1.02  $\pm$    0.02  &    1.03  $\pm$    0.03  &    1.03  $\pm$    0.03  &   \\ 
     38  &      18:53:18.72  &      +1:25:54.96  &    21.2  $\times$    10.2  &    67.2  &    0.26  &   130.5  &    17.7  $\pm$    10.8  & [   55.14  ,   61.68  ] &   58.69  $\pm$    0.03  &    1.13  $\pm$    0.02  &    0.97  $\pm$    0.02  &    0.96  $\pm$    0.02  &   \\ 
     39  &      18:53:18.70  &      +1:25:54.59  &    23.1  $\times$     6.0  &    81.3  &    0.21  &   130.5  &    11.4  $\pm$     6.6  & [   57.25  ,   60.84  ] &   58.61  $\pm$    0.03  &    0.59  $\pm$    0.02  &    0.66  $\pm$    0.02  &    0.65  $\pm$    0.02  &   \\ 
     40  &      18:53:18.51  &      +1:25:11.57  &    79.7  $\times$    29.5  &    87.4  &    0.87  &   243.6  &    18.2  $\pm$     9.8  & [   52.82  ,   62.53  ] &   57.65  $\pm$    0.01  &    1.79  $\pm$    0.01  &    1.42  $\pm$    0.01  &    1.42  $\pm$    0.01  &   \\ 

%% file: affiliations.txt

\vspace{5mm}
\noindent
Author affiliations:\\

\noindent 
$^{1}$Department of Astronomy, Yunnan University, Kunming, 650091, PR China \\
$^{2}$Indian Institute of Space Science and Technology, Thiruvananthapuram 695 547, Kerala, India\\
$^{3}$Shanghai Astronomical Observatory, Chinese Academy of Sciences, 80 Nandan Road, Shanghai 200030, Peoples Republic of China \\
$^{4}$Key Laboratory for Research in Galaxies and Cosmology, Shanghai Astronomical Observatory, Chinese Academy of Sciences, 80 Nandan Road, Shanghai 200030, Peoples Republic of China \\
$^{5}$Jet Propulsion Laboratory, California Institute of Technology, 4800 Oak Grove Drive, Pasadena, CA 91109, USA\\
$^{6}$Departamento de Astronom\'ia, Universidad de Concepci\'on, Av. Esteban Iturra s/n, Distrito Universitario, 160-C, Chile \\
$^{7}$Max-Planck-Institute for Astronomy, K\"{o}nigstuhl 17, 69117 Heidelberg, Germany \\
$^{8}$Department of Physics, P.O. box 64, FI- 00014, University of Helsinki, Finland \\
$^{9}$Kavli Institute for Astronomy and Astrophysics, Peking University, 5 Yiheyuan Road, Haidian District, Beijing 100871, People's Republic of China\\
$^{10}$Department of Astronomy, Peking University, 100871, Beijing, People's Republic of China\\
$^{11}$Departamento de Astronom\'{\i}a, Universidad de Chile, Las Condes, Santiago, Chile\\
$^{12}$Department of Astronomy, The University of Texas at Austin, 2515 Speedway, Stop C1400, Austin, TX 78712-1205, USA\\
$^{13}$Korea Astronomy and Space Science Institute, 776 Daedeokdaero, Yuseong-gu, Daejeon 34055, Republic of Korea\\
$^{14}$Indian Institute of Astrophysics, Koramangala II Block, Bangalore 560 034, India\\
$^{15}$National Astronomical Observatory of Japan, National Institutes of Natural Sciences, 2-21-1 Osawa, Mitaka, Tokyo 181-8588, Japan\\
$^{16}$Korea Astronomy and Space Science Institute, 776 Daedeokdaero, Yuseong-gu, Daejeon 34055, Republic of Korea\\
$^{17}$Satyendra Nath Bose National Centre for Basic Sciences, Block-JD, Sector-III, Salt Lake, Kolkata-700 106 \\
$^{18}$Physical Research Laboratory, Navrangpura, Ahmedabad—380 009, India \\
$^{19}$School of Physics and Astronomy, Sun Yat-sen University, 2 Daxue Road, Zhuhai, Guangdong, 519082, People's Republic of China\\
$^{20}$University of Science and Technology, Korea (UST), 217 Gajeong-ro, Yuseong-gu, Daejeon 34113, Republic of Korea\\
$^{21}$National Astronomical Observatories, Chinese Academy of Sciences, Beijing 100101, China  \\
$^{22}$SOFIA Science Centre, USRA, NASA Ames Research Centre, MS-12, N232, Moffett Field, CA 94035, USA \\

%% file: ourwork.bbl
\begin{thebibliography}{}
\makeatletter
\relax
\def\mn@urlcharsother{\let\do\@makeother \do\$\do\&\do\#\do\^\do\_\do\%\do\~}
\def\mn@doi{\begingroup\mn@urlcharsother \@ifnextchar [ {\mn@doi@}
  {\mn@doi@[]}}
\def\mn@doi@[#1]#2{\def\@tempa{#1}\ifx\@tempa\@empty \href
  {http://dx.doi.org/#2} {doi:#2}\else \href {http://dx.doi.org/#2} {#1}\fi
  \endgroup}
\def\mn@eprint#1#2{\mn@eprint@#1:#2::\@nil}
\def\mn@eprint@arXiv#1{\href {http://arxiv.org/abs/#1} {{\tt arXiv:#1}}}
\def\mn@eprint@dblp#1{\href {http://dblp.uni-trier.de/rec/bibtex/#1.xml}
  {dblp:#1}}
\def\mn@eprint@#1:#2:#3:#4\@nil{\def\@tempa {#1}\def\@tempb {#2}\def\@tempc
  {#3}\ifx \@tempc \@empty \let \@tempc \@tempb \let \@tempb \@tempa \fi \ifx
  \@tempb \@empty \def\@tempb {arXiv}\fi \@ifundefined
  {mn@eprint@\@tempb}{\@tempb:\@tempc}{\expandafter \expandafter \csname
  mn@eprint@\@tempb\endcsname \expandafter{\@tempc}}}

\makeatother

\bibitem[\protect\citeauthoryear{{\'A}lvarez-Guti{\'e}rrez et al.}{2021}]{Alv21} {\'A}lvarez-Guti{\'e}rrez R.~H., Stutz A.~M., Law C.~Y., Reissl S., Klessen R.~S., Leigh N.~W.~C., Liu H.-L., et al., 2021, ApJ, 908, 86. doi:10.3847/1538-4357/abd47c
\bibitem[\protect\citeauthoryear{Ballesteros-Paredes et al.}{2011}]{Bal11} Ballesteros-Paredes J., Hartmann L.~W., V{\'a}zquez-Semadeni E., Heitsch F., Zamora-Avil{\'e}s M.~A., 2011, MNRAS, 411, 65. doi:10.1111/j.1365-2966.2010.17657.x
\bibitem[\protect\citeauthoryear{Beuther et al.}{2018}]{Beu18} Beuther H., Mottram J.~C., Ahmadi A., Bosco F., Linz H., Henning T., Klaassen P., et al., 2018, A\&A, 617, A100. doi:10.1051/0004-6361/201833021
\bibitem[\protect\citeauthoryear{Cortes et al.}{2008}]{Cor08} Cortes P.~C., Crutcher R.~M., Shepherd D.~S., Bronfman L., 2008, ApJ, 676, 464. doi:10.1086/524355
\bibitem[\protect\citeauthoryear{Duarte-Cabral et al.}{2021}]{Dua21} Duarte-Cabral A., Colombo D., Urquhart J.~S., Ginsburg A., Russeil D., Schuller F., Anderson L.~D., et al., 2021, MNRAS, 500, 3027. doi:10.1093/mnras/staa2480
\bibitem[\protect\citeauthoryear{Evans et al.}{2021}]{Eva21} Evans N.~J., Heyer M., Miville-Desch{\^e}nes M.-A., Nguyen-Luong Q., Merello M., 2021, ApJ, 920, 126. doi:10.3847/1538-4357/ac1425
\bibitem[\protect\citeauthoryear{Evans et al.}{2020}]{Eva20} Evans N.~J., Kim K.-T., Wu J., Chao Z., Heyer M., Liu T., Nguyen-Lu'o'ng Q., et al., 2020, ApJ, 894, 103. doi:10.3847/1538-4357/ab8938
\bibitem[\protect\citeauthoryear{Gonz{\'a}lez Lobos \& Stutz}{2019}]{Gon19} Gonz{\'a}lez Lobos V., Stutz A.~M., 2019, MNRAS, 489, 4771. doi:10.1093/mnras/stz2512
\bibitem[\protect\citeauthoryear{Hacar et al.}{2013}]{Hac13} Hacar A., Tafalla M., Kauffmann J., Kov{\'a}cs A., 2013, A\&A, 554, A55. doi:10.1051/0004-6361/201220090
\bibitem[\protect\citeauthoryear{Hacar et al.}{2016}]{Hac16} Hacar A., Kainulainen J., Tafalla M., Beuther H., Alves J., 2016, A\&A, 587, A97. doi:10.1051/0004-6361/201526015
\bibitem[\protect\citeauthoryear{Hacar et al.}{2018}]{Hac18} Hacar A., Tafalla M., Forbrich J., Alves J., Meingast S., Grossschedl J., Teixeira P.~S., 2018, A\&A, 610, A77. doi:10.1051/0004-6361/201731894
\bibitem[\protect\citeauthoryear{Heyer et al.}{2009}]{Hey09} Heyer M., Krawczyk C., Duval J., Jackson J.~M., 2009, ApJ, 699, 1092. doi:10.1088/0004-637X/699/2/1092
\bibitem[\protect\citeauthoryear{Heyer \& Brunt}{2004}]{Hey04} Heyer M.~H., Brunt C.~M., 2004, ApJL, 615, L45. doi:10.1086/425978
\bibitem[\protect\citeauthoryear{Kennicutt}{2005}]{Ken05} Kennicutt R.~C., 2005, IAUS, 227, 3. doi:10.1017/S1743921305004308
\bibitem[\protect\citeauthoryear{Larson}{1981}]{Lar81} Larson R.~B., 1981, MNRAS, 194, 809. doi:10.1093/mnras/194.4.809
\bibitem[\protect\citeauthoryear{Lee et al.}{2014}]{Lee14} Lee K.~I., Fern{\'a}ndez-L{\'o}pez M., Storm S., Looney L.~W., Mundy L.~G., Segura-Cox D., Teuben P., et al., 2014, ApJ, 797, 76. doi:10.1088/0004-637X/797/2/76
\bibitem[\protect\citeauthoryear{Liu, Stutz, \& Yuan}{2019}]{Liu19} Liu H.-L., Stutz A., Yuan J.-H., 2019, MNRAS, 487, 1259. doi:10.1093/mnras/stz1340
\bibitem[\protect\citeauthoryear{Liu et al.}{2020a}]{Liu20a} Liu H.-L., Sanhueza P., Liu T., Zavagno A., Tang X.-D., Wu Y., Zhang S., 2020a, ApJ, 901, 31. doi:10.3847/1538-4357/abadfe
\bibitem[\protect\citeauthoryear{Liu et al.}{2021}]{Liu21a} Liu H.-L., Liu T., Evans N.~J., Wang K., Garay G., Qin S.-L., Li S., et al., 2021, MNRAS, 505, 2801. doi:10.1093/mnras/stab1352
\bibitem[\protect\citeauthoryear{Liu et al.}{2022}]{Liu21} Liu H.-L., Tej A., Liu T., Issac N., Saha A., Goldsmith P.~F., Wang J.-Z., et al., 2022, MNRAS, 510, 5009. doi:10.1093/mnras/stab2757
\bibitem[\protect\citeauthoryear{Liu et al.}{2020b}]{Liu20b} Liu T., Evans N.~J., Kim K.-T., Goldsmith P.~F., Liu S.-Y., Zhang Q., Tatematsu K., et al., 2020b, MNRAS, 496, 2821b. doi:10.1093/mnras/staa1501
\bibitem[\protect\citeauthoryear{Liu et al.}{2020c}]{Liu20c} Liu T., Evans N.~J., Kim K.-T., Goldsmith P.~F., Liu S.-Y., Zhang Q., Tatematsu K., et al., 2020c, MNRAS, 496, 2790c. doi:10.1093/mnras/staa1577
\bibitem[\protect\citeauthoryear{Lu et al.}{2014}]{Lu 14} Lu X., Zhang Q., Liu H.~B., Wang J., Gu Q., 2014, ApJ, 790, 84. doi:10.1088/0004-637X/790/2/84
\bibitem[\protect\citeauthoryear{Lu et al.}{2021}]{Lu 21} Lu Z.-J., Pelkonen V.-M., Juvela M., Padoan P., Haugb{\o}lle T., Nordlund {\r{A}}., 2021, arXiv, arXiv:2111.08887
\bibitem[\protect\citeauthoryear{Motte, Bontemps, \& Louvet}{2018}]{Mot18} Motte F., Bontemps S., Louvet F., 2018, ARA\&A, 56, 41. doi:10.1146/annurev-astro-091916-055235
\bibitem[\protect\citeauthoryear{Padoan et al.}{2020}]{Pad20} Padoan P., Pan L., Juvela M., Haugb{\o}lle T., Nordlund {\r{A}}., 2020, ApJ, 900, 82. doi:10.3847/1538-4357/abaa47
\bibitem[\protect\citeauthoryear{Peretto et al.}{2013}]{Per13} Peretto N., Fuller G.~A., Duarte-Cabral A., Avison A., Hennebelle P., Pineda J.~E., Andr{\'e} P., et al., 2013, A\&A, 555, A112. doi:10.1051/0004-6361/201321318
\bibitem[\protect\citeauthoryear{Rathborne et al.}{2005}]{Rat05} Rathborne J.~M., Jackson J.~M., Chambers E.~T., Simon R., Shipman R., Frieswijk W., 2005, ApJL, 630, L181. doi:10.1086/491656
\bibitem[\protect\citeauthoryear{Rathborne, Jackson, \& Simon}{2006}]{Rat06} Rathborne J.~M., Jackson J.~M., Simon R., 2006, ApJ, 641, 389. doi:10.1086/500423
\bibitem[\protect\citeauthoryear{Rosolowsky et al.}{2008}]{Ros08} Rosolowsky E.~W., Pineda J.~E., Kauffmann J., Goodman A.~A., 2008, ApJ, 679, 1338. doi:10.1086/587685
\bibitem[\protect\citeauthoryear{Shepherd, N{\"u}rnberger, \& Bronfman}{2004}]{She04} Shepherd D.~S., N{\"u}rnberger D.~E.~A., Bronfman L., 2004, ApJ, 602, 850. doi:10.1086/381050
\bibitem[\protect\citeauthoryear{Shepherd et al.}{2007}]{She07} Shepherd D.~S., Povich M.~S., Whitney B.~A., Robitaille T.~P., N{\"u}rnberger D.~E.~A., Bronfman L., Stark D.~P., et al., 2007, ApJ, 669, 464. doi:10.1086/521331
\bibitem[\protect\citeauthoryear{Shirley}{2015}]{Shi15} Shirley Y.~L., 2015, PASP, 127, 299. doi:10.1086/680342
\bibitem[\protect\citeauthoryear{Shu, Adams, \& Lizano}{1987}]{Shu87} Shu F.~H., Adams F.~C., Lizano S., 1987, ARA\&A, 25, 23. doi:10.1146/annurev.aa.25.090187.000323
\bibitem[\protect\citeauthoryear{Smith, Glover, \& Klessen}{2014}]{Smi14} Smith R.~J., Glover S.~C.~O., Klessen R.~S., 2014, MNRAS, 445, 2900. doi:10.1093/mnras/stu1915
\bibitem[\protect\citeauthoryear{Solomon et al.}{1987}]{Sol87} Solomon P.~M., Rivolo A.~R., Barrett J., Yahil A., 1987, ApJ, 319, 730. doi:10.1086/165493
\bibitem[\protect\citeauthoryear{Storm et al.}{2014}]{Sto14} Storm S., Mundy L.~G., Fern{\'a}ndez-L{\'o}pez M., Lee K.~I., Looney L.~W., Teuben P., Rosolowsky E., et al., 2014, ApJ, 794, 165. doi:10.1088/0004-637X/794/2/165
\bibitem[\protect\citeauthoryear{Tang et al.}{2019}]{Tan19} Tang Y.-W., Koch P.~M., Peretto N., Novak G., Duarte-Cabral A., Chapman N.~L., Hsieh P.-Y., et al., 2019, ApJ, 878, 10. doi:10.3847/1538-4357/ab1484
\bibitem[\protect\citeauthoryear{Urquhart et al.}{2013}]{Urq13} Urquhart J.~S., Thompson M.~A., Moore T.~J.~T., Purcell C.~R., Hoare M.~G., Schuller F., Wyrowski F., et al., 2013, MNRAS, 435, 400. doi:10.1093/mnras/stt1310
\bibitem[\protect\citeauthoryear{V{\'a}zquez-Semadeni et al.}{2019}]{Vaz19} V{\'a}zquez-Semadeni E., Palau A., Ballesteros-Paredes J., G{\'o}mez G.~C., Zamora-Avil{\'e}s M., 2019, MNRAS, 490, 3061. doi:10.1093/mnras/stz2736
\bibitem[\protect\citeauthoryear{Wang et al.}{2011}]{Wan11} Wang K., Zhang Q., Wu Y., Zhang H., 2011, ApJ, 735, 64. doi:10.1088/0004-637X/735/1/64
\bibitem[\protect\citeauthoryear{Wang et al.}{2014}]{Wan14} Wang K., Zhang Q., Testi L., van der Tak F., Wu Y., Zhang H., Pillai T., et al., 2014, MNRAS, 439, 3275. doi:10.1093/mnras/stu127
\bibitem[\protect\citeauthoryear{Yuan et al.}{2018}]{Yua18} Yuan J., Li J.-Z., Wu Y., Ellingsen S.~P., Henkel C., Wang K., Liu T., et al., 2018, ApJ, 852, 12. doi:10.3847/1538-4357/aa9d40
\bibitem[\protect\citeauthoryear{Yun et al.}{2021}]{Yun21} Yun H.-S., Lee J.-E., Evans N.~J., Offner S.~S.~R., Heyer M.~H., Cho J., Gaches B.~A.~L., et al., 2021, ApJ, 921, 31. doi:10.3847/1538-4357/ac193e
\bibitem[\protect\citeauthoryear{Zhang et al.}{2009}]{Zha09} Zhang Q., Wang Y., Pillai T., Rathborne J., 2009, ApJ, 696, 268. doi:10.1088/0004-637X/696/1/268
\bibitem[\protect\citeauthoryear{Zhang \& Wang}{2011}]{Zha11} Zhang Q., Wang K., 2011, ApJ, 733, 26. doi:10.1088/0004-637X/733/1/26







\end{thebibliography}
